\documentclass[sigconf]{acmart}

\acmConference[MSR 2022]{MSR '22: Proceedings of the 19th International Conference on Mining Software Repositories}{May 23–24, 2022}{Pittsburgh, PA, USA}

\usepackage{amsmath,amssymb,amsfonts}
\usepackage{algorithmic}
\usepackage{graphicx}
\usepackage{textcomp}
\usepackage{xcolor}
\usepackage{balance}
\usepackage{epstopdf}
\usepackage{xspace}
\usepackage{enumitem}
\usepackage{booktabs}
\usepackage{multirow}
\usepackage{tabularx}
\usepackage{color,soul}
\usepackage{hyperref}

\AtBeginDocument{
  \providecommand\BibTeX{{
    \normalfont B\kern-0.5em{\scshape i\kern-0.25em b}\kern-0.8em\TeX}}}

\acmYear{2022}
\acmDOI{10.1145/1122445.1122456}

\acmBooktitle{MSR '22: Proceedings of the 19th International Conference on Mining Software Repositories,
  May 23-24, 2022, Pittsburgh, PA, USA}
\acmPrice{15.00}
\acmISBN{978-1-4503-XXXX-X/18/06}

\begin{document}

\newcommand\mymodel{LineVD\xspace}

\title{\mymodel: Statement-level Vulnerability Detection using Graph Neural Networks}

\author{David Hin}
\affiliation{
   \institution{CREST - The Centre for Research on Engineering Software Technologies, University of Adelaide}}
\affiliation{
  \institution{Cyber Security Cooperative Research Centre}
  \city{Adelaide}
  \country{Australia, 5005}
}
\email{david.hin@adelaide.edu.au}

\author{Andrey Kan}
\affiliation{
   \institution{AWS AI Labs*\thanks{*This work was done prior to joining Amazon}}
   \city{Adelaide}
   \state{SA}
   \country{Australia, 5005}}
\email{avkan@amazon.com}

\author{Huaming Chen}
\affiliation{
   \institution{CREST - The Centre for Research on Engineering Software Technologies, University of Adelaide}}
\affiliation{
  \institution{Cyber Security Cooperative Research Centre}
  \city{Adelaide}
  \country{Australia, 5005}
}
\email{huaming.chen@adelaide.edu.au}

\author{M. Ali Babar}
\affiliation{
   \institution{CREST - The Centre for Research on Engineering Software Technologies, University of Adelaide}}
\affiliation{
  \institution{Cyber Security Cooperative Research Centre}
  \city{Adelaide}
  \country{Australia, 5005}
}
\email{ali.babar@adelaide.edu.au}

\begin{abstract}
  Current machine-learning based software vulnerability detection methods are primarily conducted at the function-level. However, a key limitation of these methods is that they do not indicate the specific lines of code contributing to vulnerabilities. This limits the ability of developers to efficiently inspect and interpret the predictions from a learnt model, which is crucial for integrating machine-learning based tools into the software development workflow. Graph-based models have shown promising performance in function-level vulnerability detection, but their capability for statement-level vulnerability detection has not been extensively explored. While interpreting function-level predictions through explainable AI is one promising direction, we herein consider the statement-level software vulnerability detection task from a fully supervised learning perspective. We propose a novel deep learning framework, \mymodel, which formulates statement-level vulnerability detection as a node classification task. \mymodel leverages control and data dependencies between statements using graph neural networks, and a transformer-based model to encode the raw source code tokens. In particular, by addressing the conflicting outputs between function-level and statement-level information, \mymodel significantly improve the prediction performance without vulnerability status for function code. We have conducted extensive experiments against a large-scale collection of real-world C/C++ vulnerabilities obtained from multiple real-world projects, and demonstrate an increase of 105\% in F1-score over the current state-of-the-art.
\end{abstract}



\maketitle
\section{Introduction}
Identifying potential software vulnerabilities is a crucial step in defending against cyber attacks \cite{cybersec}. However, it can be difficult and time-consuming for developers to determine which parts of the software will contribute to vulnerabilities in large software systems. Consequently, interest in more accurate and efficient automated software vulnerability detection (SVD) solutions has been increasing \cite{svd2012survey, svd2017survey}. Automated SVD can be broadly classified into two categories: (1) traditional methods, which can include both static and dynamic analysis, and (2) data-driven solutions, which leverage data mining and machine learning to predict the presence of software vulnerabilities \cite{svdtaxonomy}. Traditional static tools are often rule-based, leveraging knowledge from security domain experts. This can result in inconsistent performance due to a high number of false positive alerts, or completely missing more complex vulnerabilities \cite{sastempirical}. As it is challenging to define vulnerable patterns in an accurate and comprehensive way, data-driven solutions have become a promising alternative.

One primary reason for the increasing popularity of data-driven solutions can be attributed to the ever growing quantity of open-source vulnerability data. The accumulating security vulnerabilities in open-source software are regularly updated and reported to sources like the National Vulnerability Database \cite{nvd}. There have been many efforts that utilize the publicly available information to reduce the need for manually defining patterns and to learn from the data \cite{sajnani2016sourcerercc,kim2017vuddy}. As a result, data-driven solutions have generally shown outstanding performance in comparison to traditional static application security testing tools \cite{rolandsast}. One potential reason could be due to the learning ability of the models to incorporate latent information from patterns where vulnerabilities would appear in source code, in addition to the underlying causes of the vulnerability.

Despite the success of current data-driven approaches in the identification of software vulnerabilities, they are often limited to a coarse level of granularity. The model outputs often present developers with limited information for prediction outcome validation and interpretation, leading to extra efforts when evaluating and mitigating the software vulnerabilities. Consequently, many proposed SVD solutions have transitioned to either function-level \cite{devign, bgnn4vd, reveal, ivdetect} or slice-level \cite{vuldeepecker, sysevr, vuldeelocator, deepwukong} predictions, which are a major improvement from file-level predictions \cite{hovsepyan2016newer,scandariato2014predicting,du2019leopard}. Some other works further leverage supplementary information, such as commit-level code changes with accompanying log messages, to build the prediction model \cite{hoang2020cc2vec,pornprasit2021jitline}. While the goal is to help practitioners to prioritize the defective codes, vulnerabilities can often be localized to a few key lines \cite{duan2019vulsniper}. Hence, reviewing large functions could still be a considerable burden. From a preliminary analysis on a large C/C++ vulnerability detection dataset \cite{bigvul}, we find that vulnerable functions in the dataset are on average 95 lines of code.

\begin{figure}[t]
    \includegraphics[width=\linewidth]{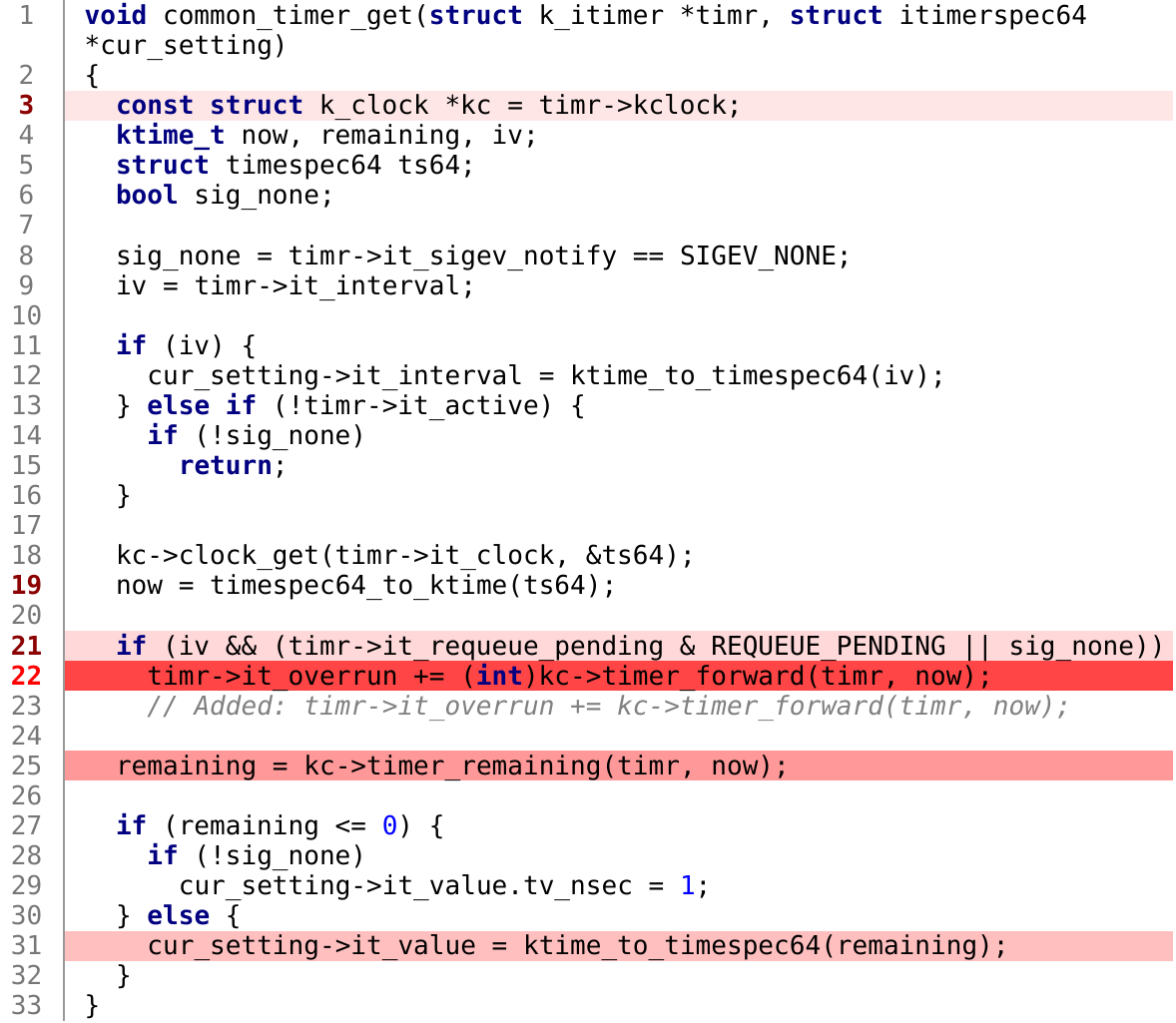}
    \caption{Vulnerable lines predicted by \mymodel for a code snippet from CVE-2018-12896, which are indicated by the presence of a red background. A darker red background color indicates higher prediction confidence by \mymodel. Ground-truth vulnerable lines have red line numbers, while dark red line numbers indicate a data or control dependency on an added line (indicated by line 23).}
    \label{predexamples}
\end{figure}

Fig.~\ref{predexamples} shows an example of how statement-level SVD can be beneficial. To save space, we choose a vulnerability from a smaller function, which contains an integer overflow vulnerability from the Linux kernel (CVE-2018-12896) that can ultimately be exploited to cause denial-of-service. With explicit statement-level predictions, it can be easier to interpret why the function has been predicted as vulnerable (or alternatively, verify that the prediction was erroneous). Focusing the developer's attention on the highlighted lines can help a developer narrow down the lines needing further inspection based on model confidence. In this case, a statement-level SVD model flags the addition assignment operation on line 22, which contains the vulnerable integer casting operation, as most suspicious, allowing a developer to more efficiently validate and mitigate the vulnerability.\par

Refining SVD granularity towards the statement-level is still in its infancy; latest work by Li et. al \cite{ivdetect} has indicated the possibility of leveraging the interpretable ML model, namely GNNExplainer, to derive the vulnerable statements as the interpretation of learnt model. However, in our work, we find that the performance is not sufficient and effective when classifying and ranking the latent vulnerable statements. Alternatively, we aim to explore the feasibility and effectiveness of \textit{directly} training and predicting on vulnerabilities at the statement level for SVD granularity refinement, which would allow data-driven solutions to directly utilize any available statement-level information in a fully supervised manner.\par

In this paper, we propose a novel framework for statement-level SVD, namely \mymodel. We focus on revisiting core components of data-driven SVD from the perspective of statement-level classification, which can serve as a way of allowing developers to better interpret vulnerability predictions. We have explored statement-level SVD in a thorough way to find the best data-driven architecture to achieve optimal performance. With the coverage of various feature extraction methods and model architectures to tackle the latent challenges in SVD, we show that \mymodel could provide sufficient capacity to incorporate contextual information for each statement in an efficient manner. The extensive evaluation of the model has been delivered in realistic scenarios; namely, heavily imbalanced labels and cross-project testing. In summary, this paper makes the following contributions:


\begin{itemize}
    \item We propose a novel and efficient statement-level SVD approach, \mymodel. With the investigation of current state-of-the-art interpretation-based SVD model showing declined performance, \mymodel achieves a significant improvement for an increase of $105\%$ in F1-score. 
    \item We investigate the performance effects of each stage for building a GNN-based statement-level SVD model, including the node embedding methods and  GNN model selection. Upon the findings, \mymodel is developed to largely improve the performance via learning from the function- and statement-level information simultaneously.
    \item \mymodel is the first approach to jointly learn from function-level and statement-level information via graph neural networks to enhance the SVD performance, which in the empirical evaluation has significantly outperformed the traditional models only using one type information.
    \item We publish our dataset, source code, and models with supporting scripts \cite{package2022}, which provides a ready-to-use implementation solution for future work with regards to benchmarking and comparison.
\end{itemize}




\section{Background}
\label{sec:background}
In this section, we will introduce relevant key concepts relating to the source code embedding and graph neural networks (GNNs), which have played key roles for providing effective and explainable capabilities to SVD prediction models in recent literature works.\par
\subsection{Source code embedding} 
To tackle a language-related task with machine learning models, it is necessary to transform the related textual corpora into vector representations. The corpora are usually specific to a given domain; e.g., one recently popular topic is source code-related tasks. The process is commonly referred to as building a language model and extracting language embeddings, which can be at the word, sentence, or document-level \cite{word2vec}. Source code modelling is the application of language modelling to source code, which can be considered a special type of structured natural language \cite{trietdeepcode}, and has demonstrated encouraging results for a wide range of downstream tasks. These include code completion \cite{liu2020multi} and code clone detection \cite{wang2020detecting}, among others \cite{wang2021mulcode}.\par
For software vulnerability detection, prior approaches have utilized document embedding methods like Doc2Vec \cite{doc2vec}, or word embedding methods such as GloVe \cite{glove} and Word2Vec \cite{word2vec} to generate pre-trained vectors for singular tokens, which are then aggregated in some way. For example, Cao et al. \cite{bgnn4vd} utilized averaged Word2Vec embeddings to transform raw code statements into vector representations.\par
Recently, transformer-based models have been applied to source code modelling, allowing for large source code understanding models to be pre-trained. They are expected to obtain higher-quality embeddings for source code, which can be leveraged for downstream tasks requiring less labeled data and training resources. One major advancement is CodeBERT \cite{codebert}, based on the RoBERTa \cite{roberta} architecture, which was trained on six different programming languages: Python, Java, JavaScript, PHP, Ruby, and Go. Specifically, CodeBERT was trained on 2.1 million natural language and programming language bimodal samples, and 6.4 million programming language unimodal samples. It should be noted that CodeBERT and Doc2Vec can produce contextual embeddings, in contrast to GloVe and Word2Vec embeddings, which are static and hence have the same embedding for each word regardless of its context.\par
Tokenization approaches for source code can vary significantly. For word embedding-based approaches, code tokens can be split by whitespace alongside punctuation, such as parenthesis and semi-colons \cite{sysevr}. Alternatively, punctuation can be completely removed \cite{ivdetect}. In addition, code identifiers are sometimes further tokenized according to common naming conventions, such as underscores or camel case. This is to help reduce the number of out-of-vocabulary tokens, which can otherwise be significantly higher than regular natural language due to the nature of how developers name identifiers. An alternative approach to manually defined tokenization rules is unsupervised subword tokenization. In CodeBERT, byte-pair encoding (BPE) \cite{bpe} is used to tokenize the source code tokens. In particular, long variable names are split into subwords based on the BPE algorithm. For example, 'add\_one' may be tokenized to 'add', '\_', and 'one'. This is arguably a more consistent way of reducing out-of-vocabulary issues in source code identifiers, as this approach is not reliant on pre-defined rules.\par

\subsection{Graph Neural Network}
\label{sec:bg_GNN}
Recently, graph neural networks have demonstrated superior performance at mining graph data structures for social networks \cite{bian2019network}, spatial-temporal related traffic networks \cite{chen2019gated}, and so on. For the downstream source code modelling tasks, transformer-based models have demonstrated promising results \cite{allamanis2018learning}. However, the complex syntactic and semantic characteristics, which are inherently presented in the programming languages, are not explicitly leveraged. \par
\begin{figure}[h]
    \includegraphics[width=0.95\linewidth]{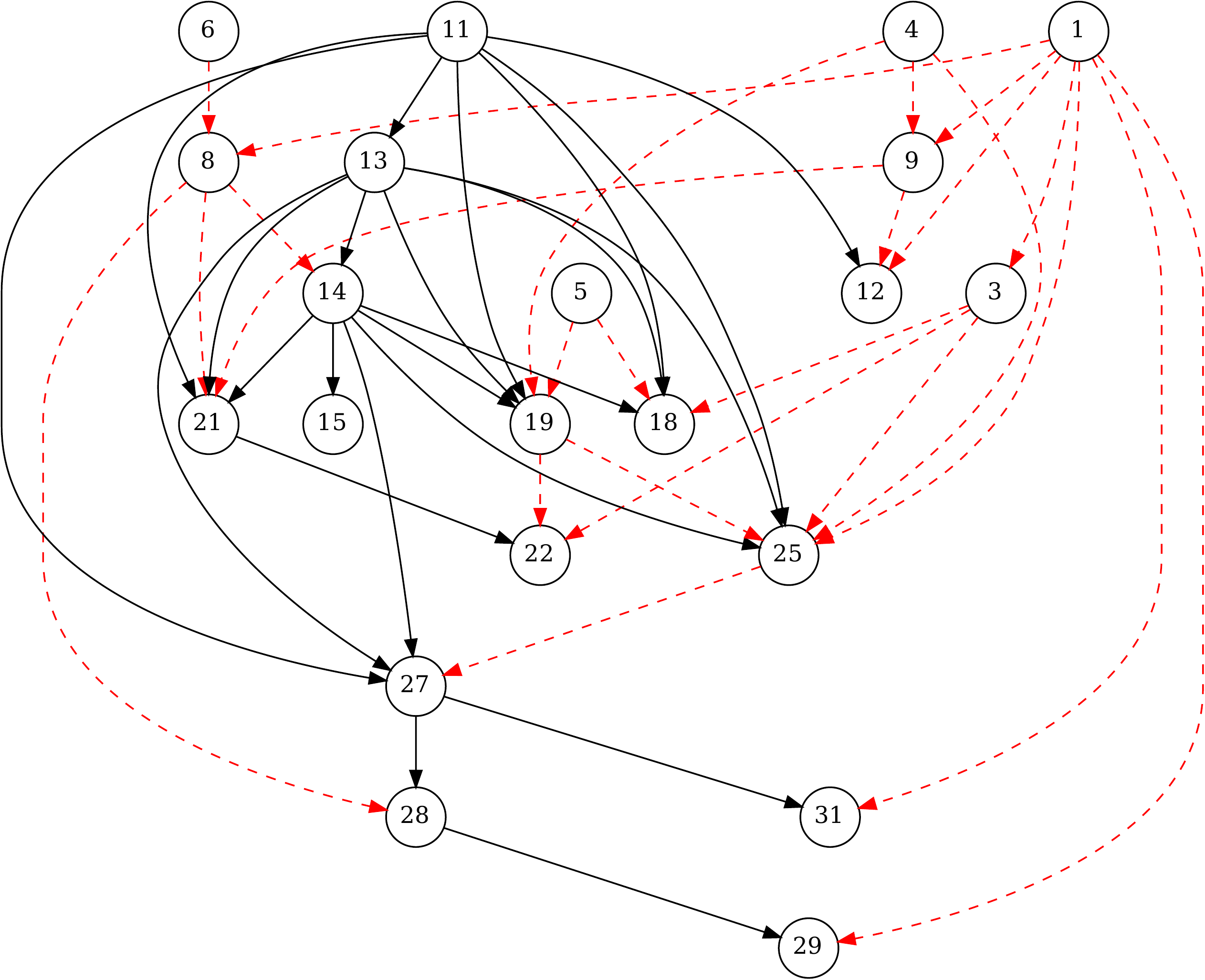}
    \caption{Program Dependence Graph for Fig. \ref{predexamples}. Black lines represent control dependency edges; dashed red lines represent data dependency edges.}
    \label{pdg}
\end{figure}

Rather than solely representing source code as a sequence of tokens, the intrinsic structure information for source code can also be effectively modelled by representing a source code snippet (or program) as a graph, $\mathcal{G} = (\mathcal{V}, \mathcal{E})$, allowing a model to more easily learn latent relationships within the source code where $\mathcal{V}$ is the set of nodes representing the program graph and $\mathcal{E}$ is the edge matrix. With different types of program-related structures, the edge matrix $\mathcal{E}$ denotes the corresponding syntactic and semantic information of the source code. Using this graph-based representation in combination with graph neural networks has resulted in improved performance for function-level vulnerability detection, in which a single graph represents a single source code function (see Section \ref{sec:related_work} for further details).
The program dependency graph (PDG) of a program (see Fig.~\ref{pdg}) is an overall focus in this work, as software vulnerabilities often involve data and control flows \cite{recurringvd,yamaguchi2014modeling,ivdetect}.
\section{The \mbox{\mymodel} Framework}

In this section, we present the \mymodel framework by firstly summarizing the problem definition. The overall architecture is subsequently discussed with details for three main components. Particularly, we demonstrate how we have incorporated the large-scale pretrained model and developed the novel graph construction method.\par

\begin{figure*}[!t]
    \centering
    \includegraphics[width=\textwidth]{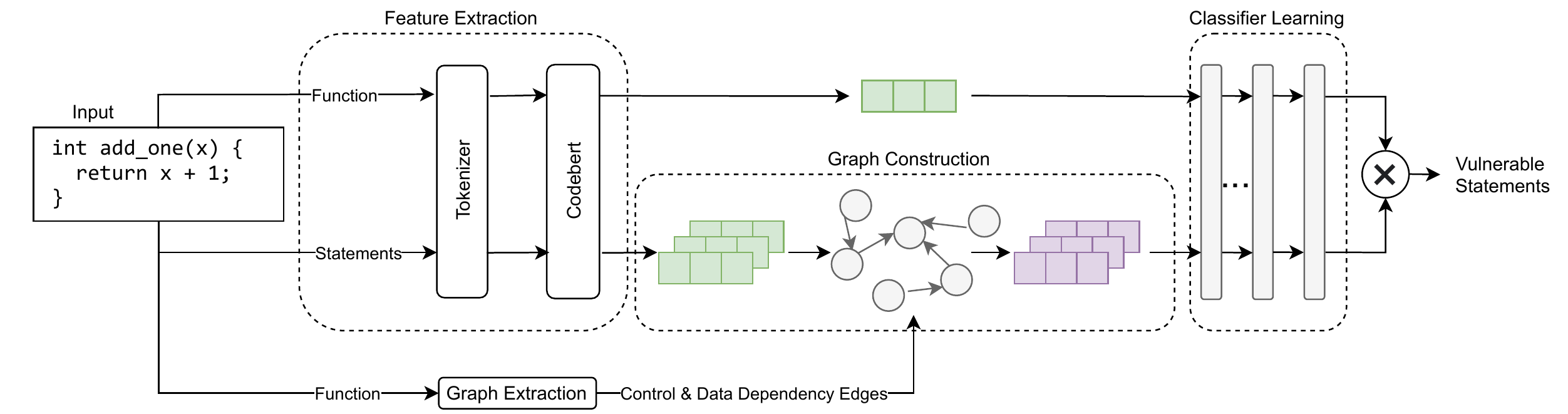}
    \caption{\mymodel Overall Architecture}
    \label{modelarch}
\end{figure*}

\subsection{Problem Definition}
\label{probdef}
We formalize the identification of vulnerable statements in a function as a binary node classification problem, i.e., learning to predict which source code statements in a function are vulnerable. Let us define a sample of data as $\left(\left(V_i, Y_i\right) | V_i \in \mathcal{V}, Y_i \in \{0,1\}, i=\{1, 2, ..., n\}\right)$, where $\mathcal{V}$ is the set of all nodes representing a statement of code in the dataset and $Y_i$ is the statement-level label where 1 is vulnerable and 0 is non-vulnerable. We collectively represent $\mathcal{Y}$ for the set of labels $Y_i$. $n$ is the number of nodes in the dataset.\par
For each $V_i$, we utilize the $n$-hop neighborhood graph $\mathcal{G}_i = (N\mathcal{V}_i, N\mathcal{E}_i, \mathcal{X}_i)$ to encode $V_i$ with the contextual information from neighboring nodes. $N\mathcal{V}_i$ indicates the neighborhood nodes for $V_i$, $N\mathcal{E}_i$ represents the corresponding edges as an adjacency matrix, $\mathcal{X}_i \in \mathbb{R}^{m\times d}$ is the node feature matrix for $V_i$, and $m$ is the number of nodes in $N\mathcal{V}_i$. The goal of \mymodel is to learn a mapping $f: \mathcal{V} \rightarrow \mathcal{Y}$ to determine the label of a given node; i.e., whether a statement is vulnerable or not. The prediction function of \mymodel, which is denoted as $f$, can be learned by minimizing the following loss function:\par
\begin{equation}
    min \sum_{i=1}^{n} \mathcal{L}({f(\mathcal{G}_i,  Y_i|V_i)})
\end{equation}
where $\mathcal{L}$ is the cross entropy loss function.

\subsection{Approach Overview}
In this section, we present a GNN-based approach to identify the vulnerable statements. One fundamental element is that the identified data and control dependencies between statements could sufficiently serve as the contextual information for the statement-level SVD task. Furthermore, we propose a novel architecture to better leverage the semantics conveyed within the statement and between the statements, which overall framework is illustrated in Figure. \ref{modelarch}. \mymodel can be divided into three main components, described in the following sections.\par

\subsubsection{Feature Extraction}
Given a snippet of source code, achieving an informative and comprehensive code representation is critical for subsequent model construction. \mymodel is firstly designed to extract the code features against a transformer-based method, which is considered as effective for source code related tasks with the self-supervised learning objectives.\par
\mymodel takes a single function of source code as the raw input. By processing and splitting the function into individual statements $V_i$, each sample is firstly tokenized via CodeBERT's pretrained BPE tokenizer. Following the collection of $V=\{V_1, V_2, ..., V_n\}$, the entire function and the individual statements comprising the function are passed into CodeBERT. Thus, the function-level and statement-level code representation can be acquired.\par
Specifically, \mymodel has separately embedded the function-level and statement-level codes, rather than aggregating the statement-level embeddings for the function-level embeddings. CodeBERT is a bimodal model, meaning it was trained on both the natural language description of a function in addition to the function code itself. As input, it uses a special separator token to distinguish the natural language description from the function code. While the natural language descriptions for the functions is not accessible, a general operation, as specified in the literature, is applied in this work to prepend each input with an additional separator token, leaving the description blank. For the output of CodeBERT, we utilize the embedding of the classification token, which is suited for code summarization tasks. This allows us to better leverage the powerful pretrained source code summarization capability of the CodeBERT model.\par
Overall, the feature extraction component of \mymodel using CodeBERT produces $n+1$ feature embeddings: one embedding for the overall function, and $n$ embeddings for each statement, for which we denote as $X^{v} = \{x^{v}_1, x^{v}_2, ..., x^{v}_n\}$ separately.\par

\subsubsection{Graph Construction}
In \mymodel, we focus on the data and control dependency information, for which we have introduced the graph attention network (GAT) model \cite{gat}. As discussed in Sec.~\ref{sec:bg_GNN}, graph neural networks (GNNs) learn the graph structured data based on an information diffusion mechanism rather than squashing the information into a flat vector, which update the node states according to the graph connectivity to preserve the important information, i.e., the topological dependency information \cite{scarselli2008graph}.\par
As shown in Figure. \ref{modelarch}, a graph attention network is used to construct the model for learning topological dependency information from the graph. A GAT layer will firstly take the $n$ output statement embeddings from CodeBERT along with the edges between each node. The graph structure, including the nodes and edges information, for the function is extracted and provided to GAT. Self-loops are added to include a given node within the set of its neighborhood graph. We initialize the GAT layer state vector $\{h^{(l)}_1, h^{(l)}_2, h^{(l)}_3,..., h^{(l)}_n\}$ with $X^{v}$. $l$ indicates the current state. \mymodel will propagate the information by embedding the data and control dependent statement (i.e., the program dependence graph) between neighboring statements in an incremental manner. Therefore, two graph attention networks are implemented in the \mymodel architecture. Overall, GAT is defined by its use of attention over features of neighbors in its aggregation function in Eq. (2) - (5):
\begin{eqnarray}
    z_i^{\left(l\right)} &=& W^{\left(l\right)}h_i^{\left(l\right)} \\
    e_{i,j}^{\left(l\right)} &=& \text{LeakyReLU}\left(\vec{a}^{\left(l\right)^T} \left(z_i^{(l)}\|z_j^{(l)}\right)\right) \\
    \alpha_{i,j}^{\left(l\right)} &=& \frac{\exp(e_{i,j}^{\left(l\right)})}{\sum_{k\in\mathcal{N}\left(i\right)}{\exp(e_{i,k}^{\left(l\right)})}} \\
    h_i^{\left(l+1\right)} &=& \sigma\left(\sum_{j\in\mathcal{N}\left(i\right)}{\alpha_{i,j}^{\left(l\right)}z_j^{\left(l\right)}}\right)
\end{eqnarray}
where $l$ is the current state, $h_i^{\left(l\right)}$ is the node embedding vectors at current layer, $W^{\left(l\right)}$ is the learnable weight matrix, $\vec{a}$ is a learnable weight vector, and $\sigma$ is an activation function.\par
\subsubsection{Classifier Learning}
As multi-layer perceptron (MLP) models are dominantly evaluated as one of the top classifiers \cite{lessmann2008benchmarking, naeem2020identifying}, we leverage the superior learning capability of a deep neural network to better train the classifier. In this work, the goal is to train a model that could jointly learn from the function-level and statement-level code simultaneously.\par
To achieve this, we consider that both function-level and statement-level code snippet will contribute equally to the prediction outcomes. Thus, we build a shared set of linear and dropout layers taking the input of function-level CodeBERT embedding and the statement embeddings obtained from the GAT layer. The ReLU~\cite{relu} function serves as the activation function. In addition, \mymodel retains the consistency of the prediction outcomes from its learning algorithm. It incorporates a element-wise multiplication between the output class of each statement and the output class of the function-level embedding, which could be either one or zero. While the vulnerable statement could be sufficient to direct the function to be vulnerable, we build \mymodel with the element-wise multiplication to further leverage the function-level information for training. Moreover, this operation harmoniously balance the conflicting outputs between function-level and statement-level embeddings, and will justify the decision for some scenarios, i.e., if the output class of the function-level embedding is zero, then all statement-level outputs are also zero. The intuition for this is that a non-vulnerable function cannot have vulnerable lines. \mymodel outputs the predictions corresponding to the statements in the input function. The cross-entropy loss function is used to train the \mymodel. More details of the implementation can be found in the replication package ~\cite{package2022}.

\section{Experimental Design and Setup}
In this section, we will report the experiment design details, including the research questions and evaluation process. Particularly, we present the dataset details for the empirical evaluation. We further discuss the applied evaluation metrics, which is considered to thoroughly quantify the model performance.
\subsection{Research Questions}
To explore statement-level vulnerability detection task and investigate the performance of \mymodel, we answer and motivate the following research questions:

\textbf{RQ1: How much performance increasement can \mymodel achieve in comparison with the state-of-the-art interpretati-on-based SVD model?}
To evaluate the relative improvement of \mymodel in the statement-level SVD task, we choose to compare against the state-of-the-art interpretation-based model. It is conducted from two diverse measures, which are binary classification and ranked metrics for vulnerable statements.

\textbf{RQ2: How do different code embedding methods affect statement-level vulnerability detection?}
Code embedding methods have not yet been explored for statement-level SVD compared to SVD at other levels of granularity.

\textbf{RQ3: How do graph neural networks and function-level information contribute to \mymodel performance?}
The effect of information propagation using graph neural networks on statement-level SVD has yet to be explored.

\textbf{RQ4: How does \mymodel perform in a cross-project classification scenario?}
While training on a dataset containing multiple projects already reduces misrepresentation of model generalisability, it is still possible for samples from the same project to appear in both the training and test set. Using a cross-project scenario can better represent how the model performs on a completely unseen project, rather than only unseen samples.

\textbf{RQ5: Which statement types are best distinguished by Lin-eVD for real-world data?}
Investigating the model prediction outcomes from the perspective of statement types, particularly for real-world data, can help to understand where the model performs best and where it fails, which can guide future work and improvements in statement-level SVD.

\subsection{Datasets}
Recent research suggests that SVD models should be evaluated on data that could represent the distinct characteristics of real-world vulnerabilities \cite{reveal}. This means evaluating on source code extracted from real-world projects (i.e. non-synthetic) while maintaining an imbalanced ratio, which is inherent for vulnerabilities in software projects. The usage of datasets that do not satisfy these conditions would result in the inconsistency of model performance when applied in real world scenarios. Another dataset requirement is a sufficiently large number of samples, ideally spanning multiple projects in order to acquire a model that can generalize well to unseen code. The final requirement is access to ground truth labels at the statement level or traceability to the before-fix code, i.e., the original git commit.\par
By extracting the code vulnerabilities from over 300 different open-source C/C++ GitHub projects, Big-Vul contains the trustworthy source code vulnerabilities spanning 91 different vulnerability types, which are linked to the public Common Vulnerabilities and Exposures (CVE) database~\cite{bigvul}. A substantial amount of manual resources has also been dedicated to ensure the quality of the dataset. Meanwhile, it has provided the enriched information including CVE IDs, CVE severity scores and particularly the code changes, along with other metadata. In the end, Big-Vul provides the best fit for meeting the requirements to model code-centric vulnerability detection to the best of our knowledge, containing approximately 10,000 vulnerable samples and 177,000 non-vulnerable samples. This large diversity in projects, rather than focusing on a limited number of projects in particular, allows for better representation of all open-source C/C++ projects (see Table \ref{rq4} for the top 10 most common projects in the dataset).


\subsubsection{Ground-truth Labels} 
To obtain the ground-truth labels for the vulnerable and non-vulnerable lines, we follow the assertions from literature \cite{ivdetect, bigvul} rather than proposing our own heuristics: (1) removed lines in a vulnerability-fixing commit serve as an indicator for a vulnerable line, and (2) all lines that are control or data dependent on the added lines are also treated as vulnerable. The reasoning for the second point is that any added lines in a vulnerability-fixing commit were added to help patch the vulnerability. Hence, lines that were not modified in vulnerability-fixing commit, but are related to these added lines, can be considered as related to the vulnerability.


To obtain labels corresponding to lines that are dependent on the added lines, we first obtain the code changes from the before and after version of the sample, where a sample in Big-Vul refers to a function-level code snippet. For the before version, we remove all the added lines, and for the after version, we remove all the deleted lines. In both cases, we keep blank placeholder lines to ensure line number consistency. The code graph extracted from the after version can be used to find all lines that are control or data dependent on the added lines, whose line numbers correspond to the before version. This set of lines can be combined with the set of deleted lines to obtain the final set of vulnerable lines for a single sample. Commented lines are excluded in the code graph, and hence are not used for training or prediction. This can be seen in Fig. \ref{predexamples} and \ref{pdg}; in this case, there is only a modified line, which is treated as both a deleted line (22) and added line (23). The control and data dependency edges in this case happen to be the same for both the before and after version, and hence we can use Fig. \ref{pdg} to identify the lines that are control/data dependent on line 23, which are lines 3, 19, and 21.

\subsubsection{Dataset Cleaning} 
We performed multiple filtering steps on the Big-Vul dataset. First, we remove all comments from the code. Second, we ignore code changes that are purely cosmetic, such as changes to whitespace, and consequently remove any functions with no non-cosmetic code changes. Third, we removed improperly truncated functions. A few samples in the original dataset were truncated incorrectly, resulting in an unparsable and invalid code sample. For example, a function that was originally 50 lines may be incorrectly truncated to 40 lines for no apparent reason. The reason may be due to an error in how the dataset was originally constructed; however, there were only 30 such samples in the whole dataset. We use a random training/validation/test split ratio of 80:10:10. For the training set, we undersample the number of non-vulnerable samples to produce an approximately balanced dataset at the function-level, while the test and validation set is left in the original imbalanced ratio. We choose to balance the samples at the function-level as it is non-trivial to balance at the statement-level while maintaining the contextual dependencies between statements within a function.

\subsection{Evaluation Metrics}
We report F1 score, precision, recall, area under the receiver operating characteristic curve (ROCAUC), and area under the precision-recall curve (PRAUC). While ROCAUC is widely used to directly measure the predictive power of the model without choosing a specific threshold, it cannot fully reflect the effectiveness when dealing with imbalanced datasets. Hence, in addition to reporting ROCAUC for comparison with past literature, we have also used PR-AUC, which is better suited for imbalanced problems. \par 
While binary classification is the primary focus, as described in Section \ref{probdef}, we also report ranked metrics to evaluate the performance of the most confident predictions of the model. The ranked metrics include mean average precision (MAP), normalized discounted cumulative gain (nDCG), and mean first ranking (MFR). Here, we define first ranking as the rank of the first correctly predicted vulnerable statement. We care about how well the model performs at a certain number of $k$ most confident lines, as we can thus further limit the amount of code that needs to be reviewed by the developer. In this work, we choose to use $k=5$ as an effective performance representation at the top following the recommendation of Li et al. \cite{ivdetect}. However, in practical usage, this threshold could be adjusted by the user. Finally, we report accuracy of interpretation given N nodes, where N is equal to 5 (denoted as N5 in Table \ref{stmt_rank_res}). This can be interpreted as function-level accuracy given only the prediction results of the top five lines.\par 
Due to the imbalanced nature of the dataset, we first find the best threshold for the F1-score using the validation set before calculating the F1-score on the test set. When determining the significance of improvements, we use the Wilcoxon signed-rank test \cite{wilcoxon} on the F1-scores of ten runs with random seeds. The best models according to the automated hyper parameter tuning are used for the test.\par

\subsection{Hyper parameters}
Hyper parameters of \mymodel were tuned using Ray Tune \cite{raytune} with randomized grid search. The hyper parameter details can be found in the publicly released source code. The scores are reported corresponding to the mean test results across ten runs using the best hyper parameters based on the loss of the validation set.

\subsection{Experimental Design of Research Questions}

\subsubsection{RQ1}
In RQ1, we focus on the comparison between our proposed model and the existing literature methods. For SVD task, a practical experiment setting is that the vulnerability remains unknown for a given piece of code regardless at function level or file level. Thus, leveraging the interpretable ML models for the validation and interpretation of prediction results of SVD models is dominant. In this work, we compare \mymodel with the state-of-the-art interpretation-based SVD model, namely \textit{IVDetect} from ~\cite{ivdetect}. \textit{IVDetect} is designed as an interpretable vulnerability detector which embeds both artificial intelligence to detect vulnerabilities and intelligence assistant to provide vulnerabilities interpretations from statements. It has presented the state-of-the-art performance in terms of the detection and localization capabilities with the graph-based software vulnerability detection and interpretation models. While it has included an empirical evaluation against the existing function-level deep learning-based approaches, a full replication of \textit{IVDetect} for comparison in RQ1 is provided in terms of the performance of vulnerable statements identification.

\subsubsection{RQ2} 
\label{rq2subsection}
In RQ2, we investigate the impacts of utilizing pretrained embedding methods for statement-level vulnerability prediction. While vectorizing the code statement information (also node information in this work) is a critical step, we have included different feature embedding methods, including CodeBERT, Doc2Vec, and averaged GloVe embeddings to understand their impacts on prediction performance. These two baselines were chosen to represent the embedding methods usually seen in vulnerability detection models \cite{sysevr, ivdetect, vuldeepecker, deepwukong, bgnn4vd}. GloVe and Word2Vec embeddings usually perform similarly in the vulnerability detection context \cite{glovew2vvd}, and thus we only use GloVe, which was also the chosen word embedding technique used in IVDetect \cite{ivdetect}. For classification of the embeddings, we use multiple hidden layers, and freeze all parameters of the CodeBERT model during training.


\subsubsection{RQ3}
In RQ3, we explore the effect of introducing GNN layers into the feature extraction component of the model. In this experiment, we compare the use of two popular GNN variants: Graph Convolution Networks (GCN) \cite{gcn} and Graph Attention Networks (GAT) \cite{gat}. These two GNN types were chosen to explore the effect of different GNN architectures on statement-level vulnerability classification. The GNN layers are inserted after the feature extraction component and before the final hidden layers. We also test two different types of graphs extracted from the source code: Program Dependence Graphs (PDG) and Control Dependence Graphs (CDG). PDGs consist of both data dependency edges and control dependency edges. Data dependency edges describe how data flows between statements in the program (which statements are influenced by which variables), while control dependency edges describe the order in which statements execute, as well as whether they execute or not. We use both variants to explore whether GCNs and GATs can utilize data dependency information from related statements to better distinguish vulnerable statements. The control and data dependencies are extracted using the Joern program \cite{yamaguchi2014modeling}. We remove samples that either cannot be parsed by Joern, or are correctly parsed but do not contain control or data dependency edges. Since the nodes produced by Joern are not necessarily at the line-level, we group together nodes with the same line number, and remove nodes with no line numbers (e.g. metadata nodes). \par
In addition to the GNN component, we simultaneously explore how to incorporate the function-level information, for which it may benefit \mymodel to reduce the false positive rate of statement-level classification. This is done through training on the function-level label of the function, which is embedded using CodeBERT.

\subsubsection{RQ4}
In RQ4, we explore how \mymodel performs in a cross-project scenario; i.e. samples from the target project do not appear in the source projects which are used in model training. We simulate this by producing multiple splits where the test set is made of a single project, and the rest of the projects are used in the training and validation set. We choose the top 10 projects with the most vulnerable samples in the dataset to use as cross-project splits.

\subsubsection{RQ5}
In RQ5, we examine which statement types (e.g. if-statement, goto-statement) \mymodel can correctly distinguish. We use the node types given by Joern. For the "Control Structure" node type, we replace it with the given control structure type (e.g. if, while, for). For "Function Call" node types, we split them into two different categories: "built-in" function calls, which are functions that appear in the C standard library, and "external" function calls, which are any other functions not found in the standard C library. This is an important distinction, as each sample in the dataset consists of a code snippet at the function-level. This means the only information present in external function calls is in the identifier name itself, along with its arguments, rather than the contents of the external function. For "Operator" node types, we group them into the following categories: assignment, arithmetic, comparisons, access, and logical. Any operator node types that do not fall into these categories are grouped into "other".\par

\section{Results}
We run our experiments on a computing cluster utilizing multiple NVIDIA Tesla V100 GPUs and Xeon E5-2698v3 CPUs operating at 2.30 GHz.
\subsection{RQ1: How much performance advantage can \mymodel achieve in comparison with the state-of-the-art SVD model?}

Table \ref{stmt_rank_res} and Table \ref{stmt_clf_res} summarize the performance comparison of \mymodel with respect to IVDetect, a recently proposed fine-grained vulnerability detection approach, using various evaluation measures. We show that \mymodel significantly outperforms IVDetect method in all metrics. In Table \ref{stmt_rank_res}, the ranked metrics are included to show the accuracy of different models. A higher accuracy means that the model can generate more precise vulnerability detection at the statement level. As seen in Table \ref{stmt_rank_res}, \mymodel largely outperforms IVDetect in all ranked metrics ($p<0.01$). For ranked metrics, \mymodel improves $N5$ by $0.205$ in accuracy value from $0.695$ to $0.900$, and increases MAP@5 by $0.336$ over IVDetect from $0.424$ to $0.760$. For MFR, \mymodel improves the performance by $4.373$ ranks, which significantly increase the efficiency of correctly locating the first vulnerable statement by 114.5\%. \par
\begin{figure}
    \includegraphics[width=\linewidth]{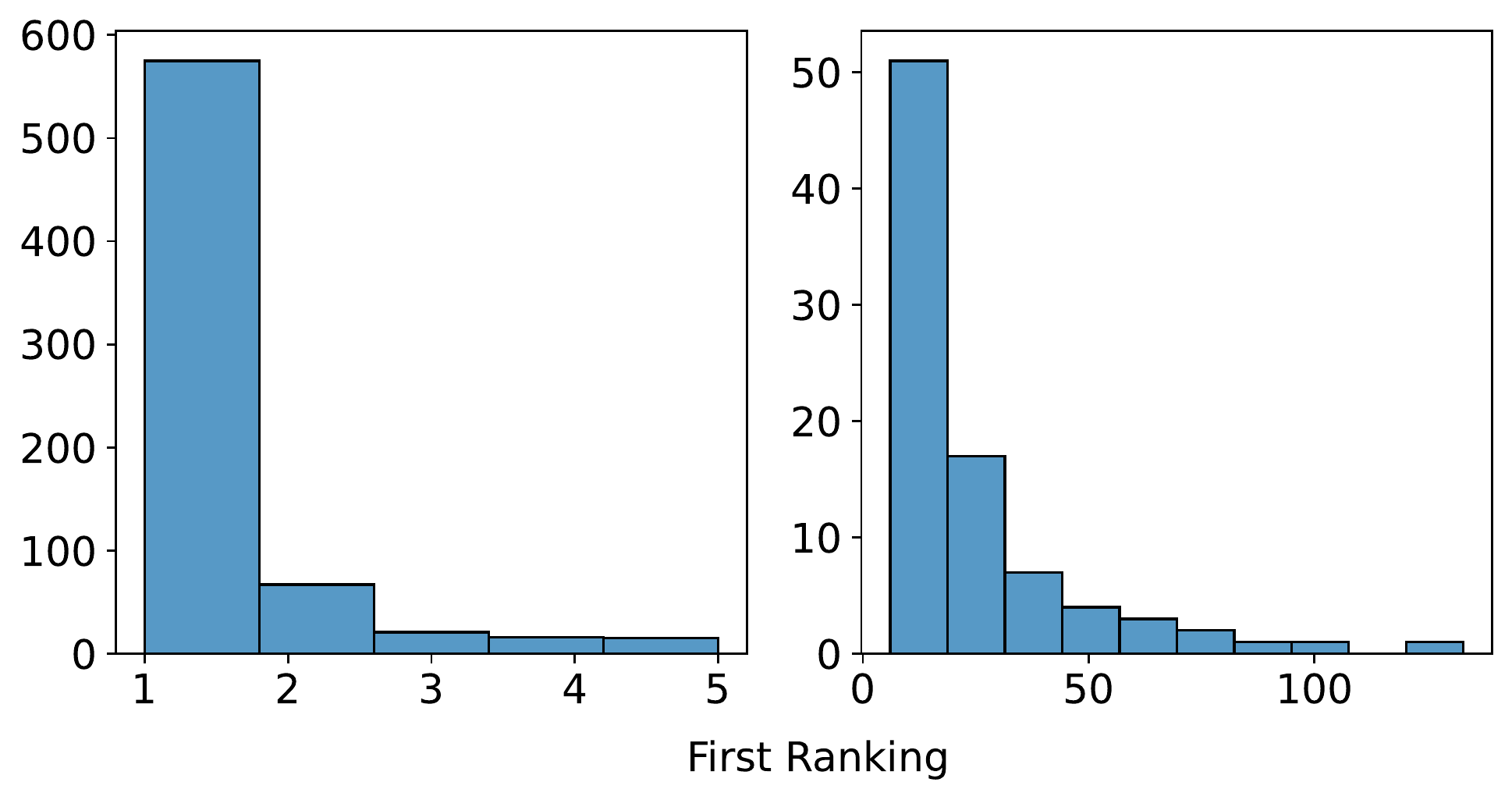}
    \caption{Histogram of first rankings of \mymodel on default test set. First ranking is defined as the first true-positive statement in a sorted list of softmax scores assigned to each statement for a given function. E.g. in 575 samples, a vulnerable statement is first in the ranked list of vulnerable statement predictions; in 67 samples, a vulnerable statement is second in the list, etc.}
    \label{firsts}
\end{figure}

When examining the distribution of rankings for \mymodel, we find that it is extremely skewed towards the lower end, with some highly incorrect predictions raising the MFR. The distribution of first-rank scores is plotted in Figure. \ref{firsts}, which shows that the majority (89\%) of first ranking scores are between 1 to 5. This suggests that \mymodel may struggle with certain types of longer functions, but correctly ranks them in most cases.\par
Table \ref{stmt_clf_res} provides the binary classification metrics, which is the most straightforward evaluation (whether or not a line is correctly predicted as vulnerable statement), and the one directly aligned with our problem definition. Overall, \mymodel outperforms IVDetect by 104\% ($p<0.01$) in F1-score. Specifically, for practical usage, we notice that recall score has also been substantially improved from $0.140$ to $0.533$, indicating that \mymodel has boosted the ability to correctly determine and locate the vulnerable statements.\par
We also note that another advantage of our architecture regards its efficiency in generating statement-level predictions compared to IVDetect. The use of an explanation model significantly increases the inference time for a single sample, as an entire model must be trained to find a subgraph that explains the model's prediction, compared to the single forward pass required by \mymodel. Using an NVIDIA Tesla V100, a single inference using GNNExplainer can take over a minute, depending on its configuration, while the single forward pass in \mymodel takes less than a second.

\begin{table}[!b]
    \caption{RQ1: Statement-level Performance (Ranked)}
    \label{stmt_rank_res}
    \centering
    \begin{tabularx}{\columnwidth}{l|XXXXX}
        \toprule
        Methods  & N5             & MAP@5          & NDCG@5         & MFR            \\
        \midrule
        IVDetect & 0.695          & 0.424          & 0.517          & 8.192          \\
        \mymodel & \textbf{0.900} & \textbf{0.760} & \textbf{0.804} & \textbf{3.819} \\
        \bottomrule
    \end{tabularx}
\end{table}

\begin{table}[!b]
    \caption{RQ1: Statement-level Performance (classification)}
    \label{stmt_clf_res}
    \centering
    \begin{tabular*}{\columnwidth}{l|@{\extracolsep{\stretch{1}}}*{5}{l}}
        \toprule
        Methods                      & F1             & Rec            & Prec           & ROCAUC         & PRAUC          \\
        \midrule
        IVDetect                     & 0.176          & 0.140          & 0.238          & 0.463          & 0.520          \\
        \mymodel                     & \textbf{0.360} & \textbf{0.533} & \textbf{0.271} & \textbf{0.913} & \textbf{0.642} \\
        \bottomrule
    \end{tabular*}
\end{table}


\subsection{RQ2: How do different code embedding methods affect statement-level vulnerability detection?}
We evaluate the effectiveness of CodeBERT regarding the code embedding methods. To be able to justify and simplify the comparison with same process, as stated in section.~\ref{rq2subsection}, the multi-layer perceptron model is chosen and we limit the embedding process for statement-level information. In Table \ref{rq2}, CodeBERT has outperformed the baseline feature embedding methods, which are Doc2Vec and GloVe, in the context of statement-level vulnerability prediction. CodeBERT increases the value of F1-score in comparison to Doc2Vec by 134\% ($p<0.01$), and GloVe by 16\% ($p<0.05$) respectively. This is to be expected, as the pre-trained CodeBERT model is designed with over 125 million parameters, which provide an enhanced capability to encode richer information in a larger and deeper model from code snippets comparing to the other models with less layers and parameters. \par 
We also note that, while CodeBERT has the advantage over GloVe and Doc2Vec being pre-trained on a large corpus, it has the disadvantage of being trained on an external dataset consisting of non C/C++ code in five other programming languages. In comparison, the GloVe and Doc2Vec models are directly trained on our C/C++ dataset. Since Doc2Vec should be able to provide a better encoding capability for the contextual information about statements, GloVe presents a second best performance by the averaged aggregation method. Nonetheless, CodeBERT provides a best code embedding method given the comparative experiments with the other most popular code embedding methods for SVD to date.
\begin{table}[hbt]
    \caption{RQ2: Feature Embedding Methods for Statement-level Vulnerability Classification}
    \label{rq2}
    \centering
    \begin{tabularx}{\columnwidth}{l|XXXXX}
        \toprule
        Embedding & F1             & Rec            & Prec           & ROCAUC         & PRAUC          \\
        \midrule
        Doc2Vec   & 0.064          & 0.167          & 0.040          & 0.580          & 0.508          \\
        GloVe     & 0.129          & 0.166          & 0.106          & 0.666          & 0.529          \\
        CodeBERT  & \textbf{0.150} & \textbf{0.254} & \textbf{0.121} & \textbf{0.703} & \textbf{0.534} \\
        \bottomrule
    \end{tabularx}
\end{table}

\subsection{RQ3: Can graph neural networks and function-level information benefit statement-level classification?}

Table. \ref{table:rq3} shows the effect of different experiment settings involving GNN type, program graph type, and whether or not the function-level classification component is included in the model. As it could be seen from Table. \ref{table:rq3}, using graph attention network for the feature learning from PDG information is the best fit, which we have hence included as the graph component in \mymodel. When comparing this GNN feature extraction combination with the model variation without GNN, we achieve an increase of 24\% in F1 score ($p<0.01$) from 0.296 to 0.360, indicating that the presence of the GNN benefits the model's learning capability. However, the performance of other graph-based combinations is generally comparable to using the model without a GNN. This suggests that the program graph type and GNN type non-trivially affects the performance of the model. In particular, only using control dependency edges generally results in lower performance than using both control and data dependencies, suggesting that data dependency edges are important in predicting statement-level vulnerabilities.\par 
In addition, we find that GCN achieves worse performance in comparison to GAT for all model types, which is to be expected, as the feature aggregation in GCN treats all neighboring nodes equally, unlike GAT, which attends to certain neighbors. However, when comparing only using statement-level information for classification (i.e. no function-level information involved), the use of GCN with either of the graph types or a GAT with CDG results in worse performance compared to using no GNN. Finally, we find that the enrichment of function-level information in the model significantly increases the statement-level performance, regardless of whether a GNN is used. Using the GAT+PDG combination, we achieve a performance increase of 140\% ($p<0.01$).

\begin{table}[hbt]
    \caption{RQ3: Graph-based Feature Variants for Statement-level Vulnerability Classification}
    \label{table:rq3}
    \centering
    \begin{tabularx}{1.02\columnwidth}{l|ccccc}
        \toprule
        Model Type                & F1             & Rec            & Prec           & ROCAUC         & PRAUC          \\
        \midrule
        GAT+CDG                   & 0.115          & 0.120          & 0.112          & 0.657          & 0.528          \\
        GAT+CDG+Func          & 0.304          & 0.491          & 0.221          & 0.907          & 0.624          \\
        GAT+PDG                   & 0.150          & 0.254          & 0.121          & 0.703          & 0.534          \\
        \textbf{GAT+PDG+Func} & \textbf{0.360} & 0.533          & \textbf{0.271} & 0.913          & \textbf{0.642} \\
        GCN+CDG                   & 0.084          & 0.143          & 0.060          & 0.632          & 0.514          \\
        GCN+CDG+Func          & 0.283          & \textbf{0.558} & 0.190          & 0.911          & 0.597          \\
        GCN+PDG                   & 0.085          & 0.125          & 0.067          & 0.599          & 0.513          \\
        GCN+PDG+Func          & 0.310          & 0.460          & 0.235          & 0.905          & 0.616          \\
        No GNN                    & 0.129          & 0.166          & 0.106          & 0.666          & 0.529          \\
        No GNN+Func           & 0.296          & 0.537          & 0.205          & \textbf{0.921} & 0.619          \\
        \bottomrule
    \end{tabularx}
\end{table}
\subsection{RQ4: How does \mymodel perform in a cross-project classification scenario?}
\begin{table}[!t]
    \caption{RQ4: Cross-project Statement-level Prediction}
    \label{rq4}
    \centering
    \begin{tabularx}{1.02\columnwidth}{l|llllll}
        \toprule
        Project     & \small{F1} & \small{Rec} & \small{Prec}  & \small{ROCAUC} & \small{PRAUC} & \small{Vuln} \\
        \midrule
        Chromium    & 0.298 & 0.470 & 0.219 & 0.923  & 0.625 & 3103 \\
        Linux       & 0.301 & 0.502 & 0.216 & 0.925  & 0.630 & 1847 \\
        Android     & 0.290 & 0.494 & 0.208 & 0.922  & 0.625 & 962  \\
        ImageMagick & 0.333 & 0.504 & 0.249 & 0.925  & 0.644 & 331  \\
        PHP         & 0.290 & 0.487 & 0.208 & 0.928  & 0.622 & 200  \\
        TCPDump     & 0.284 & 0.452 & 0.207 & 0.925  & 0.622 & 197  \\
        OpenSSL     & 0.298 & 0.508 & 0.211 & 0.926  & 0.628 & 157  \\
        Krb5        & 0.259 & 0.535 & 0.186 & 0.903  & 0.605 & 139  \\
        QEMU        & 0.250 & 0.478 & 0.177 & 0.910  & 0.601 & 120  \\
        FFmpeg      & 0.269 & 0.483 & 0.193 & 0.917  & 0.612 & 115  \\
        \bottomrule
    \end{tabularx}
\end{table}

From Table \ref{rq4}, we can see that performance is generally consistent throughout the different projects. Due to the nature of the various projects, there are varying numbers of vulnerable samples for each project. The top 10 software projects are reported in Table \ref{rq4}. While the performance varies, in general, they are all slightly lower than the results from the random splits in Table \ref{stmt_clf_res}. This is par for the course, as even partial inclusion of within-project information such as variable and function names may assist the model in distinguishing vulnerable samples.\par
In Table \ref{rq4}, the investigated projects are ranked according to the number of vulnerabilities, which is indicated in the last column `\textit{Vuln}'. We note that the Chromium and Linux splits contain a disproportionately large number of vulnerable samples. The Chromium split itself accounts for 30\% of all vulnerable samples in the dataset. Despite this, the performance is competitive with the other splits, indicating that the model attains a high generalisability for different settings, which has shown comparable performance with a smaller training set in Table \ref{rq4}.

\subsection{RQ5: Which statement types are best distinguished by \mymodel?}

Table \ref{statement_empirical} shows the raw prediction results for different statement types. We report the confusion matrix values for each statement type and sort them by F1 score. We see that that assignment operations and external function calls are the most commonly occurring vulnerable lines, and that the model most often correctly classifies Function Declaration statements. This could be due to the significantly higher level of information in function declarations compared to other statements, which can consist of the return type, function name, parameter types, and the parameter names. Operation-related statements generally perform similarly, except for logical operations, which \mymodel can generally better distinguish, and access and cast operations, which are achieve relatively poorer performance.

\mymodel has higher performance with built-in function call statements compared to external function calls. This is somewhat expected, as the same built-in functions are likely to appear across more samples, unlike user-defined function names. In addition, due to the function-level nature of the dataset, the only information in external function call statements is from the function name itself and the arguments passed. Without knowledge of what occurs within an external function call, it can be difficult to find distinguishing patterns relating to their vulnerability.

\begin{table}[!t]
    \caption{RQ5: Analysis of Statement-level Predictions}
    \label{statement_empirical}
    \centering
    \begin{tabularx}{\columnwidth}{l|XXXXXX}
        \toprule
        Statement Type         & TP   & FP   & TN    & FN   & F1   \\
        \midrule
        Function Declaration   & 530  & 364  & 17572 & 126  & 0.68 \\
        While Statement        & 48   & 72   & 1374  & 23   & 0.50 \\
        Builtin Function Call  & 142  & 246  & 4537  & 72   & 0.47 \\
        Logical Operation      & 90   & 159  & 4601  & 60   & 0.45 \\
        Switch Statement       & 18   & 31   & 1424  & 31   & 0.37 \\
        For Statement          & 75   & 195  & 3632  & 88   & 0.35 \\
        If Statement           & 749  & 1930 & 50969 & 855  & 0.35 \\
        Assignment Operation   & 1206 & 3792 & 81490 & 1051 & 0.33 \\
        Other Operation        & 62   & 200  & 5401  & 54   & 0.33 \\
        Jump Target            & 18   & 53   & 11119 & 19   & 0.33 \\
        Arithmetic Operation   & 27   & 106  & 1379  & 10   & 0.32 \\
        Return Statement       & 166  & 526  & 27332 & 186  & 0.32 \\
        External Function Call & 644  & 2212 & 63378 & 582  & 0.32 \\
        Comparison Operation   & 19   & 64   & 1489  & 17   & 0.32 \\
        Access Operation       & 44   & 168  & 3488  & 48   & 0.29 \\
        Cast Operation         & 25   & 115  & 3146  & 28   & 0.26 \\
        Continue               & 2    & 13   & 1039  & 1    & 0.22 \\
        Break                  & 5    & 42   & 7966  & 43   & 0.11 \\
        Goto Statement         & 3    & 15   & 5115  & 48   & 0.09 \\
        \bottomrule
    \end{tabularx}
\end{table}
\mymodel struggles most with continue, break, and goto statements. These are all control structure nodes that affect the control dependency graph, and do not contain any direct statement-level information that could distinguish them from each other. Hence, to distinguish the vulnerability of these statements, the model must rely on the patterns relating to the context in which they appear. Despite using a GNN to capture these relationships, these statements still cannot be classified correctly. Hence, while the use of GNN does improve overall performance, these results indicate that neither GCN or GAT are sufficient for effectively distinguishing these particular control statements. In contrast, statement types like function declarations, function calls, and for/while/if control structures often contain significantly more information within the statement, and hence can be distinguished more easily. This focus on more complex control structure statement types and built-in function calls is suitable for practical usage as they are more likely to be directly involved in the vulnerability.

\section{Discussion}
In this section, we discuss the threats to validity and limitations of our work, along with ideas for future work.

\subsection{Threats to Validity}

The first threat relates to the potential sub-optimal hyper parameter tuning; we rely on a random grid search for determining the combinations of hyper parameters; however, it would be close to impossible to exhaustively test all combinations of hyper parameters. To mitigate this threat, we adopt best practices where possible when choosing the default range of values for the hyper parameter search, and tune each model variation with the same number of random parameter configurations. 

Another threat is about the dataset. We use a maintained C/C++ dataset built for vulnerability detection as the same usage as in \cite{ivdetect}. Thus, it has appeared that the information from the dataset generally target on intral-procedural information. Most samples in current vulnerability detection datasets \cite{devign,reveal,bigvul} lack inter-procedural information (including functions from other files), which make it difficult for both learning-based approaches and regular static analysis tools to determine the underlying nature of a vulnerability. Certain requirements would be necessary, such as handling macros and inter-procedural function calls, to capture all necessary information to learn the underlying nature of the vulnerabilities. These requirements are non-trivial such that training a learning-based model is an extremely challenging task if only looking at source code. However, future works could explore the possibility of producing datasets with deeper contextual information for individual samples, which could then be used to more comprehensively test the capabilities of GNNs for solving these issues.

There has also been recent work on the effect of time-based validation on software defect prediction \cite{timesplitsdefect}, which could also be relevant to vulnerability detection. We mitigate this threat by also exploring cross-project vulnerability prediction, since time-based validation primarily applies to within-project prediction. Furthermore, any graph-based model should intuitively have similar capabilities in regards to learning patterns associated with vulnerabilities, and hence the performance loss (or gain) should be linear across the different baseline models. However, this is something that could be tested more thoroughly in future works.

\subsection{Limitations and Future Work}

While \mymodel outperforms the current state-of-the-art, the performance is still quite low, meaning there is still much room for improvement in the statement-level SVD task. A particular area of interest is finding the best way to produce better code embeddings. While CodeBERT can perform well, it is not trained on C/C++. The use of a large language model pretrained on the specific target language (e.g. C-BERT \cite{cbert}) could improve the performance of downstream tasks. However, at the time of writing, there were no openly released C/C++-based large pretrained models.

Another limitation regards the learning capability of the GNN layer, which inherently only propagates information from immediately neighboring nodes (i.e. 1-hop). To propagate further range information, multiple graph neural network layers can be stacked. For example, two GNN layers would propagate information across a 2-hop neighborhood for each node. However, most graph neural network architectures using more than just a few layers has been shown to result in over-squashing, where the exponentially growing information between each layer cannot be captured within a fixed-length vector \cite{gnnbottleneck}, resulting in bottlenecked performance. This is consistent with our findings; we tuned the number GNN layers during experimentation but did not find any significant increase in performance beyond two layers. 

Future work could explore novel GNN architectures that can better capture vulnerability patterns in program graphs. Moreover, it is recognized that, amid the two different categories of software metrics-based \cite{hovsepyan2016newer,scandariato2014predicting,du2019leopard} and pattern-based \cite{li2016vulpecker,vuldeepecker,vuldeelocator} approaches as data-driven SVD solutions, a few of them could have the potential of delivering a fine-grained prediction outcome. In the future work, the investigation of a systematic comparison with other tools would give broader and deeper perspective of evaluation.

\section{Related work}
\label{sec:related_work}
To recap the state-of-the-art of software vulnerability detection researches, we introduce the related work from two perspectives: 1) the application of GNN on SVD, 2) the explanability of machine learning models for SVD.
\subsection{Software vulnerability detection with GNN}
Detecting software vulnerability is an ongoing topic to secure software systems from cyber attacks. Recent advances have advocated the application of GNN and its variants for function level vulnerability detections, which is considered as the best way of representing source code in the SVD context \cite{plbart, devign}, and have demonstrated enhanced performance over other approaches \cite{bgnn4vd, ivdetect,duan2019vulsniper}.\par
\indent The introduction of GNNs for modelling vulnerabilities was originally inspired by the vulnerability discovery approach proposed by Yamaguchi et al. \cite{yamaguchi2014modeling} using Code Property Graphs, a type of program graph incorporating program dependency edges \cite{liu2006gplag,johnson2015exploring}, control flow edges, and the abstract syntax tree of the program, which provide an additional source of information to learn from \cite{cui2020vuldetector}. Hence, the performance improvements using GNNs can primarily be attributed to leveraging the domain knowledge that lines of source code within a program have specific relationships to other lines; i.e., training using both semantic and syntactical information, rather than only syntactical information.\par
\indent To achieve this goal, the lines of source code, which is also the nodes in the graph, will be firstly vectorized according to the possible code tokens. The vectorized information is then concatenated as initial node information to allow GNNs to capture the semantic and syntactical information. This propagation of information between semantically relevant statements theoretically allows the model to better make use of relevant contextual lines. While GNNs have been shown to perform well in function-level vulnerability classification (graph classification) \cite{devign, bgnn4vd,reveal, deepwukong, lin2021deep}, the effects of vectorized methods and GNNs models on statement-level vulnerability classification (node classification) have yet to be explored.
\subsection{Interpretation machine-learning based models for SVD}
One way to improve the performance of machine-learning based method for SVD in practice is the development of explainable detection results, which could provide a fine-grained vulnerability prediction outcome. Particularly, there have been works attempting to detecting lines-level information by leveraging explainable artificial intelligence for software engineering tasks, such as detecting source code lines for defect prediction \cite{wattanakriengkrai2020predicting,pornprasit2022deeplinedp}. This raises the importance of research for interpretable machine-learning based models. Besides the novel machine-learning based model for function level vulnerability detection output, the existing works are limited to providing partial information for the explanation generation, i.e., tokens from ~\cite{svdheuristicxai} and intermediate code by ~\cite{vuldeelocator}. Other way for statement-level SVD task may be via localizing the specific vulnerable statements with the assumption of receiving vulnerable source codes in function level~\cite{ding2022velvet}. However, it demonstrates a certain limitation as SV detection and localization are mostly demanded at the same time.\par
\indent A recently explored benefit of GNNs for SVD is direct access to explainable GNN approaches \cite{ivdetect}. These can be attached to any GNN-based function-level SVD model to obtain fine-grained predictions; i.e., statement-level predictions if the nodes themselves represent statements. However, whether or not it is the best way to build a statement-level vulnerability classifier has yet to be explored. For example, one disadvantage of using a function-level detector as the base model for interpretation is the inability to directly leverage any statement-level information during the training process, in addition to the significantly longer inference times.
\section{Conclusion}
We introduce \mymodel, a novel deep learning approach for statement-level vulnerability detection, which can allow developers to more efficiently evaluate potentially vulnerable functions. \mymodel achieves a new state-of-the-art on statement-level vulnerability detection on real-world open source projects by leveraging graph neural networks and statement-level information during training. The significant improvement in comparison to the latest fine-grained machine-learning based model indicates the effectiveness of directly utilizing statement-level information for statement-level SVD. Finally, \mymodel achieves reasonable cross-project performance, indicating its effectiveness and generalization capabilities even for completely unseen software projects. Future directions will include exploring alternate pretrained feature embedding methods and novel GNN architectures that can better accommodate the underlying nature of software source code and the vulnerabilities.

\section*{Acknowledgment}
The work was supported by the Cyber Security Research Centre Limited whose activities are partially funded by the Australian Government’s Cooperative Research Centres Programme. This work was supported with supercomputing resources provided by the Phoenix HPC service at the University of Adelaide.

\bibliographystyle{ACM-Reference-Format}
\balance
\bibliography{IEEEfull}

\end{document}